\documentclass[conference]{IEEEtran}
\usepackage[pdftex]{graphicx}
\usepackage{amsmath}
\usepackage{amssymb}
\usepackage{cite}
\usepackage{color}
\usepackage{wasysym}
\usepackage{algorithm,algpseudocode,algorithmicx}
\makeatletter
\newcommand{\algmargin}{\the\ALG@thistlm}
\makeatother
\newlength{\whilewidth}
\algdef{SE}[parWHILE]{parWhile}{EndparWhile}[1]
  {\parbox[t]{\dimexpr\linewidth-\algmargin}{%
     \hangindent\whilewidth\strut\algorithmicwhile\ #1\ \algorithmicdo\strut}}{\algorithmicend\ \algorithmicwhile}%
\algnewcommand{\parState}[1]{\State%
  \parbox[t]{\dimexpr\linewidth-\algmargin}{\strut #1\strut}}

\newlength{\indexwidth}
\algnewcommand{\parNumState}[2]{\State{#1}%
  \parbox[t]{\dimexpr\linewidth-\algmargin-\dimexpr \indexwidth}{\strut #2\strut}}

\algdef{SE}[SUBALG]{Indent}{EndIndent}{}{\algorithmicend\ }%
\algtext*{Indent}
\algtext*{EndIndent}

\algnewcommand\Input{\item[{\textbf{Input:}}]}
\algnewcommand\Output{\item[{\textbf{Output:}}]}

\IEEEoverridecommandlockouts
\begin{document}
\title{Distributed Multi-Step Power Scheduling and Cost Allocation for Cooperative Microgrids}

\author{\IEEEauthorblockN{Lu An, Jie Duan, Yuan Zhang,  Mo-Yuen Chow, Alexandra Duel-Hallen}
\thanks{This work was partially supported by the U.S. National Science Foundation under award EEC-0812121 and award CNS-1505633.}
\IEEEauthorblockA{Department of Electrical and Computer Engineering\\
North Carolina State University\\
Raleigh, NC 27695\\
Email: lan4@ncsu.edu, jduan3@ncsu.edu, yzhang50@ncsu.edu, chow@ncsu.edu, sasha@ncsu.edu}
}

\maketitle
\thispagestyle{plain}
\pagestyle{plain}

\begin{abstract}
Microgrids are self-sufficient small-scale power grid systems that can employ renewable generation sources and energy storage devices and can connect to the main grid or operate in a stand-alone mode. Most research on energy-storage management in microgrids does not take into account the dynamic nature of the problem and the need for fully-distributed, multi-step scheduling. First, we address these requirements by extending our previously proposed \textit{multi-step cooperative distributed energy scheduling} (CoDES) algorithm to include both purchasing power from and selling the surplus power to the main grid. Second, we model the microgrid as a multi-agent system where the agents (e.g. households) act as players in a cooperative game and employ a distributed algorithm based on the Nash Bargaining Solution (NBS) to fairly allocate the costs of cooperative power management (computed using CoDES) among themselves. The dependency of the day-ahead power schedule and the costs on system parameters, e.g., the price schedule and the user activity level (measured by whether it owns storage and renewable generation devices), is analyzed for a three-agent microgrid example. 
\end{abstract}


\IEEEpeerreviewmaketitle

\section{Introduction}
Microgrids support a flexible, reliable, and efficient integration of renewable sources of energy, such as solar and wind, energy storage devices, and demand response \cite{5357331,el2014smart}. Distributed scheduling of microgrid power to optimize the overall cost without utilizing a control center is vital to successful microgrid management \cite{6980137}. Most distributed power scheduling methods in the literature on microgrids assume single-step optimization and do not incorporate dynamic evolution of the stored energy \cite{5357331,el2014smart,6980137,wang2014integrating,6279592, 6466429}. In \cite{7286376,7445949}, we proposed a \textit{dynamic multi-step cooperative distributed energy scheduling} (CoDES) algorithm where the energy storage devices in the microgrid are scheduled to cooperatively \textit{minimize the overall cost} of purchasing power from the main grid. In this paper, the CoDES method \cite{7286376} is extended to include both purchasing power from and selling power to the main grid.

The extended CoDES algorithm minimizes the overall, or \textit{social}, microgrid cost. To determine individual users' costs, we model a microgrid as a dynamic \textit{multi-agent} system where each agent (e.g., a household) can own distributed energy storage devices (DESDs) and/or renewables and is associated with an individual load profile. The agents act as players in a \textit{cooperative game} \cite{6279592} and employ a computationally efficient \textit{Nash Bargaining Solution} (NBS) \cite{Avrachenkov2015265} to  fairly allocate the costs of power provided by the grid. Game theory was utilized for microgrid energy storage optimization in e.g., \cite{wang2014integrating,6279592,6466429} while cost allocation for power industry was addressed in \cite{fiestras2011cooperative, beeler2014network, hao2015distributed} as well as our recent work \cite{7525415,lian2016game}. However, the distribution of costs of microgrid power management was not addressed previously. 

In the proposed NBS cost allocation algorithm, the optimal \textit{social cost} is found first using the extended CoDES method. Then this total cost is distributed among the users according to each user's need for power provided by the main grid as well as for cooperation. The proposed cost allocation method reduces the users' costs relative to stand-alone power scheduling, thus enticing them to join in cooperative power scheduling. A novel \textit{consensus}-based \cite{4118472} method is employed to achieve the \textit{distributed} cost allocation for all agents. Finally, numerical results for a three-agent grid-connected microgrid illustrate the effects of the variation of the power prices over the 24-hour period and individual agents' loads and resources on the day-ahead optimal power schedule and the agents' costs. 

The main \textit{contributions} of this paper are:

\begin{itemize}
\item Development of a fully-distributed multi-step power scheduling method that optimizes both purchasing power from and selling the surplus power to the main grid.
\item Design of a distributed fair cost allocation method for multi-agent microgrids.
\item Validation of practical relevance and computation efficiency of proposed algorithms using a microgrid model with heterogeneous loads and activity levels and time-varying power prices.
\end{itemize}

The rest of the paper is organized as follows. Section II presents the microgrid model and the distributed CoDES method. In Section III, distributed NBS-based cost allocation is derived for multi-agent microgrids. Section IV contains numerical results and analysis, and Section V concludes the paper.

\section{The Microgrid Model and Coopeartive energy Scheduling} \label{CPS}
\begin{figure}[!t]
\centering
\includegraphics[width=0.48\textwidth]{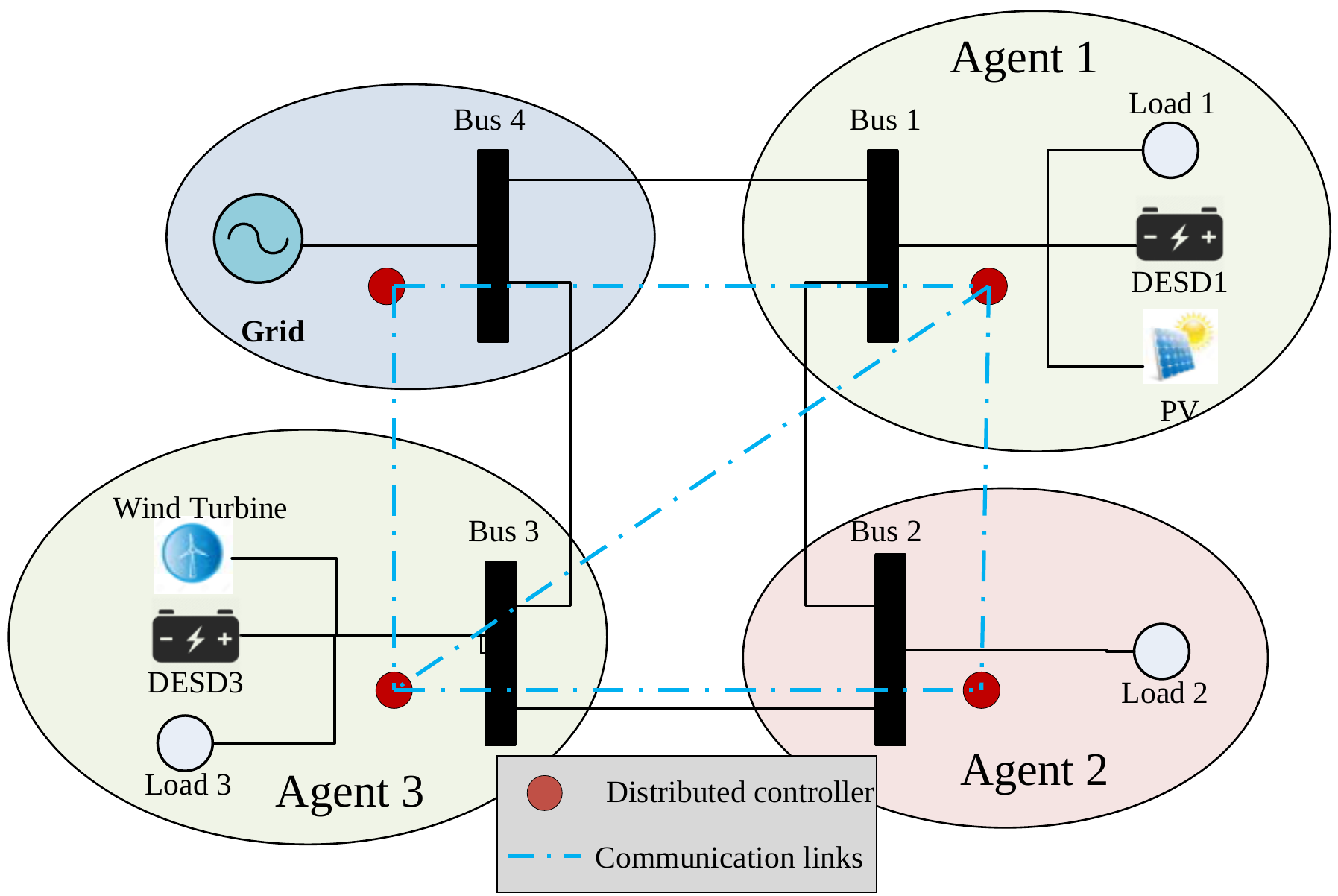}
\vspace{-0.1in}
\caption{An example of $3$-user grid-connected microgrid system model}
\label{fig:sym}
\vspace{-0.25in}
\end{figure}

Consider an $n$-Bus microgrid system where the supply side (Bus $n$) represents the main grid, and the demand side has $r=n-1$ users, including both \textit{passive} users, which own loads, and \textit{active} users, which also own DESDs and/or renewable energy generation units, e.g., rooftop PV panels or wind turbines. The passive users are the energy consumers, i.e. they can only purchase energy from the grid, while active users can also sell their generated power to the main grid. Fig. \ref{fig:sym} shows an example of a 4-Bus system, where Bus $1$ and Bus $3$ belong to active users, and Bus $2$ belongs to a passive user. Table \ref{tab:notation} includes the notation frequently used in the paper.

\begin{table}[h!]
\vspace{-0.1in}
\caption{Notation table}
\vspace{-0.1in}
\centering
\begin{tabular}{|c|p{6cm}|}
\hline
\textbf{Term} & \textbf{Definition}  \\
\hline
$U_{act}$/$U_{pas}$ & The sets of active/passive users\\
\hline
$T$ & Number of time steps for power scheduling\\
\hline
$t$ & Current time step\\
\hline
$\Delta t$ & Time interval between two time steps (hr)\\
\hline
$p_b(t)$/$p_s(t)$ & Prices at which the users purchase/sell power from/to the grid \\
\hline
$P_G^+(t) \geq 0$ & Power purchased from the grid\\
\hline
$P_G^-(t) \geq 0$ & Power sold to the grid\\
\hline
$P_G(t)$ & ${P_G}\left( t \right) =P_G^ + \left( t \right) - P_G^ - \left( t \right)$ \\
\hline
$P_G^{max}$ & The maximum power $P_G^+(t)$/$P_G^-(t)$ that can be purchased from or sold to the grid at any time step\\
\hline
$P_{i,D}(t)$ & The forecasted load demand power of the $i^{\mathrm{th}}$ user\\
\hline
$P_{i,R}(t)$ & The renewable generation power of the $i^{\mathrm{th}}$ user\\
\hline
$P_{i,B}(t)$ & The charging/discharging power command of the $i^{\mathrm{th}}$ user's DESD\\
\hline
${\mathbf{P_i}(t)}$ & $[P_G^ + (t) \quad P_G^ - (t) \quad P_{i,B}(t)]'$ is the vector of primal variables of user $i$, $i=1,\cdots ,r$\\
\hline
$P_{i,B}^{min}$/$P_{i,B}^{max}$ & The charging and discharging lower/upper limits of the $i^{\mathrm{th}}$ user's DESD power $P_{i,B}(t)$\\
\hline
$E_{i,B}(t)$ & The energy stored in the $i^{\mathrm{th}}$ user's DESD\\
\hline
$E_{i,B}^{0}$ & The initial energy stored in the $i^{\mathrm{th}}$ user's DESD\\
\hline
$E_{i,B}^{min}$/$E_{i,B}^{max}$ & The lower/upper capacity limits of the $i^{\mathrm{th}}$ user's DESD energy $E_{i,B}(t)$\\
\hline
$J$ & The total bill\\
\hline
$\lambda(t)$ & The incremental cost\\
\hline
$\Delta P(t)$ & The global power imbalance\\
\hline
$\omega_{ij}$ & The communication connectivity strength between nodes $i$ and $j$\\
\hline
$N_i$ & The set of neighbors of node $i$\\
\hline
\end{tabular}
\label{tab:notation}
\vspace{-0.1in}
\end{table}

\subsection{Optimal Power Scheduling Problem}

All demand-side users collaborate to minimize the total electricity bill for the upcoming time interval of $T$ time steps:
\begin{equation}
\label{eq:min_cost_linear}
\mathop J = \min \limits_{{{\mathbf{P}}_{\mathbf{i}}(t), \forall i}} \sum\limits_{t = 1}^T {\left( {{p_b}\left( t \right)P_G^ + \left( t \right) - {p_s}\left( t \right)P_G^ - \left( t \right)} \right)\Delta t}.
\end{equation}
In (\ref{eq:min_cost_linear}), the cost is minimized by optimizing the microgrid power schedule, which is subject to:\\
1) \textit{Power Balance Constraint}: At any time step $t$, the amount of load is equal to the amount of power generation:
\begin{equation}
\vspace{-0.1in}
\label{eq:Power_balance}
\resizebox{.91\hsize}{!}{$
\sum\limits_{i = 1}^{n} {{P_{i,D}}\left( t \right)}  = \sum\limits_{i = 1}^{n} {\left( {{P_{i,B}}\left( t \right) + {P_{i,R}}\left( t \right)} \right)}  + \left( {P_G^ + \left( t \right) - P_G^ - \left( t \right)} \right). $}
\vspace{+0.1in}
\end{equation}
where $P_G^ + \left( t \right) = 0$ or $P_G^ - \left( t \right)=0$ for $\forall t$.
\\
2) \textit{DESD Dynamics and Capacity Limits}:
The system states are given by the values of the stored energy in DESDs ${E_{i,B}}(t)$, which satisfy
\begin{equation}
\label{eq:Ba_next}
{E_{i,B}}(t + 1) = {E_{i,B}}(t) - {P_{i,B}}(t)\Delta t,\forall i \in {U_{act}}.
\end{equation}
From (\ref{eq:Ba_next}):
\begin{equation}
\label{eq:Ba_t}
E_{i,B}(t)=E_{i,B}^0-\sum\nolimits_{\tau  = 1}^t {{P_{i,B}}\left( \tau  \right)} \Delta t,\forall i \in {U_{act}}.
\end{equation}
Thus, for any $1\leq t \leq T$ and any $ i \in {U_{act}}$:
\begin{equation}
\vspace{-0.1in}
\label{eq:Ba_cons}
{E_{i,B}^0} - E_{i,B}^{\max } \leqslant \sum\nolimits_{\tau  = 1}^t {{P_{i,B}}\left( \tau  \right)} \Delta t \leqslant {E_{i,B}^0} - E_{i,B}^{\min }.
\end{equation}
\\

\subsection{Cooperative Distributed Energy Scheduling Algorithm} \label{CoDES}
First, we formulate the augmented Lagrangian function for the optimization (\ref{eq:min_cost_linear}) under the constraints (\ref{eq:Power_balance}), (\ref{eq:Ba_cons}) and Table \ref{tab:notation}:
\begin{equation}
\label{eq:aug_Lag}
\resizebox{.92\hsize}{!}{$
\begin{gathered}
  L = \sum\limits_{t = 1}^T {\left( {{p_{b}}\left( t \right)P_G^ + \left( t \right) - {p_{s}}\left( t \right)P_G^ - \left( t \right)} \right)\Delta t} 
   + \sum\limits_{t = 1}^T {\lambda \left( t \right)\Delta P\left( t \right)}  \hfill \\ + \sum\limits_{i \in {U_{act}}}^{} {\sum\limits_{t = 1}^T {{\mu _{1i}}\left( t \right)\Delta {P_{1i}}\left( t \right)} }  + \sum\limits_{i \in {U_{act}}}^{} {\sum\limits_{t = 1}^T {{\mu _{2i}}\left( t \right)\Delta {P_{2i}}\left( t \right)} }  \hfill \\
   + \frac{\rho }{2}\sum\limits_{t = 1}^T {\Delta P{{\left( t \right)}^2}}  + \frac{\rho }{2}\sum\limits_{i \in {U_{act}}} {\sum\limits_{t = 1}^T {\Delta {P_{1i}}{{\left( t \right)}^2}} }  + \frac{\rho }{2}\sum\limits_{i \in {U_{act}}} {\sum\limits_{t = 1}^T {\Delta {P_{2i}}{{\left( t \right)}^2}} } , \hfill \\ 
\end{gathered} $}
\end{equation}
where $\lambda(t)$, $\mu_{1i}(t)$ and $\mu_{2i}(t)$ are the Karush-Kuhn-Tucker (KKT) multipliers \cite{7286376}, $\rho$ is the penalty factor, and
\begin{equation}
\vspace{-0.1in}
\label{eq:delta_P}
\begin{gathered}
\Delta P\left( t \right) = \sum\nolimits_{i = 1}^{n} {{P_{i,D}}\left( t \right)} - \left( {P_G^ + \left( t \right) - P_G^ - \left( t \right)} \right)  \hfill \\ 
 - \sum\nolimits_{i \in {U_{act}}}^{} {\left( {{P_{i,B}}\left( t \right) + {P_{i,R}}\left( t \right)} \right)},
\end{gathered}
\end{equation}
\begin{equation}
\label{eq:delta_P1}
\Delta {P_{1i}}\left( t \right) = {\left[ {E_{i,B}^0 - E_{i,B}^{\max } - \sum\nolimits_{\tau  = 1}^t {{P_{i,B}}\left( \tau  \right)} \Delta t} \right]^ + },
\end{equation}
\begin{equation}
\label{eq:delta_P2}
\Delta {P_{2i}}\left( t \right) = {\left[ {E_{i,B}^{\min } - {E_{i,B}^0} + \sum\nolimits_{\tau  = 1}^t {{P_{i,B}}\left( \tau  \right)} \Delta t} \right]^ + },
\end{equation}
where $[]^+$ projects its argument to its positive values.

By taking the gradient of $L$ in (\ref{eq:aug_Lag}) with respect to the \textit{primal variables'} vector ${\mathbf{P_i}(t)} = [P_G^ + (t) \quad P_G^ - (t) \quad P_{i,B}(t)]'$ and the \textit{dual variables'} vector ${\mathbf{\Lambda_i}}(t) = [\lambda(t) \quad \mu _{1i}(t) \quad \mu _{2i}(t)]'$, we obtain the update equations:
\begin{equation}
\begin{gathered}
\label{eq:P_up}
{{\mathbf{P}}_{\mathbf{i}}}^{k + 1}(t) = {{\mathbf{P}}_{\mathbf{i}}}^k(t) - {{\boldsymbol{\xi} }_{\mathbf{1}}}{\nabla _{{{\mathbf{P}}_{\mathbf{i}}}^k(t)}}L,
\end{gathered}
\end{equation}
\vspace{-0.2in}
\begin{equation}
\begin{gathered}
\label{eq:lambda_up}
{{\mathbf{\Lambda }}_{\mathbf{i}}}^{k + 1}(t) = {{\mathbf{\Lambda }}_{\mathbf{i}}}^k(t) - {{{\boldsymbol{\xi }}}_{\mathbf{2}}}{\nabla _{{{\mathbf{\Lambda }}_{\mathbf{i}}}^k(t)}}L,
\end{gathered}
\end{equation}
where ${{\boldsymbol{\xi }}_{\mathbf{1}}}$, ${{\boldsymbol{\xi }}_{\mathbf{2}}}$ are the vectors of update coefficients.
 
However, the equations (\ref{eq:P_up}), (\ref{eq:lambda_up}) require the global parameters $\lambda(t)$ and $\Delta P(t)$, respectively, which is not available locally to individual users. To make the computataion fully distributed, the consensus algorithm is employed in each Bus to compute the estimates of the global information $\widehat \lambda_i \left( t \right)$ and $ \Delta \widehat P_i(t)$ \cite{7286376,4118472,yang2013consensus}:
\begin{equation}
\label{eq:lambda_est}
\resizebox{.88\hsize}{!}{$
\widehat \lambda _i^{k + 1}\left( t \right) = \widehat \lambda _i^k\left( t \right) + \sum\nolimits_{j \in {N_i}} {{\omega _{ij}}\left( {\widehat \lambda _j^k\left( t \right) - \widehat \lambda _i^k\left( t \right)} \right)}  + {\xi _3}\Delta \widehat P_i^k\left( t \right),$}
\end{equation}
\begin{equation}
\label{eq:Pi_est}
\resizebox{.78\hsize}{!}{$
\begin{gathered}
\Delta \widehat P_i^{k + 1}\left( t \right) = \Delta \widehat P_i^k\left( t \right) + \Delta P_i^{k + 1}\left( t \right) - \Delta P_i^k\left( t \right)  \hfill \\ + \sum\nolimits_{j \in {N_i}} {{\omega _{ij}}\left( {\Delta \widehat P_j^k\left( t \right) - \Delta \widehat P_i^k\left( t \right)} \right)}, 
\end{gathered} $}
\end{equation}
where $\omega_{ij}$ is the entry $(i,j)$ of a doubly-stochastic consensus update matrix $\mathbf{W}$, which represents the communication connectivity strength \cite{4118472} between the nodes $i$ and $j$, $N_i$ denotes the set of neighbors of node $i$, and the local power imbalance
\begin{equation}
\label{eq:Pi}
\resizebox{.88\hsize}{!}{$
\Delta P_i^{k+1}\left( t \right) = \left\{ {\begin{array}{*{5}{c}}
   {{P_{i,D}}\left( t \right),i \in {U_{pas}}} \\   
{{P_{i,D}}\left( t \right) -  P_{i,B}^k\left( t \right) - {P_{i,R}}\left( t \right) ,i \in {U_{act}}} \\ 
  { - \left( {P_G^{ + (k)}\left( t \right) - P_G^{ - (k)}\left( t \right)} \right),i \in Grid} 
\end{array}} \right.. $}
\end{equation}

By replacing $\lambda(t)$ and $\Delta P(t)$ with their estimates $\widehat \lambda_i \left( t \right)$ and $ \Delta \widehat P_i(t)$, respectively, we obtain a fully-distributed algorithm, which computes the optimal solution of the energy scheduling problem and the optimal total cost $J$ (\ref{eq:min_cost_linear}) of the microgrid $T$ time steps ahead.

\section{The Multi-Agent Microgrid Model and Cost Allocation} \label{NBS}
The solution to (\ref{eq:min_cost_linear}) optimizes the total cost of the system when all
users cooperate on microgrid power scheduling. However, it is necessary
to allocate these costs among the users. One possible cost distribution
is based on the cost of consumption, found by computing the total cost
of the power bought/sold by each agent in the optimization (\ref{eq:min_cost_linear}). However,
such costs do not represent a \textit{fair} cost allocation since they are
chosen to minimize the social objective (\ref{eq:min_cost_linear}), which does not necessarily
reflect individual users' power needs.

In this section, the microgrid is modeled as an $r$-agent system connected to the main grid as illustrated in Fig. \ref{fig:sym}. Several approaches to fair cost allocation for cooperative games have
been proposed in the literature \cite{Avrachenkov2015265,kawamori2016nash,saad2009distributed,myerson1980conference}. We employ Nash Bargaining Solution (NBS) due to its computational efficiency \cite{Avrachenkov2015265}. The proposed cost
allocation is unique and fair as discussed below.
\subsection{Overview of the Nash Barganing Solution (NBS)}
Consider a system with $r$ players and a system-wide cost function $J$. The NBS cost allocation algorithm proceeds in \textit{three steps} \cite{Avrachenkov2015265}:

(1) The players cooperate to minimize the \textit{social cost} $J$.

(2) The \textit{disagreement point} is computed as $\mathbf{D}=(D_1,D_2,\cdots, D_{r})$, where the \textit{selfish cost} $D_i$ is the maximum cost the $i$th player is willing to pay.

(3) The overall cost $J$ is split among the players, with the \textit{allocated cost} of player $i$ given by Theorem 2 in \cite{Avrachenkov2015265}:
\begin{equation}
\label{eq:Ji}
{J_i} = {D_i} - \frac{{\sum\nolimits_{i = 1}^{r} {{D_i}}}  - J}{r} \quad \forall i = 1, \cdots ,r.
\end{equation}
Note that bargaining is successful when the social cost is no greater than the sum of the selfish costs, i.e.,
\begin{equation}
\label{eq:JlessD}
J\leq \sum\nolimits_{i = 1}^r {{D_i}},
\end{equation}
or equivalently, each player's allocated cost does not exceed its selfish cost: 
\begin{equation}
\label{eq:eps}
\epsilon = D_i - J_i \geq 0,
\end{equation}
where $ \epsilon$ is the discount of cooperation, which is the same for all players. Finally, note that by Theorem 2 in \cite{Avrachenkov2015265}, the cost $J_i$ is the smallest achievable allocated cost of player $i$ given $J$ and $\mathbf{D}$ in steps 1 and 2, respectively, thus resulting in a fair cost allocation.

\subsection{Cost allocation for Multi-Agent Microgrids} \label{cl}
Next, NBS cost allocation is applied to the cooperative power scheduling problem in Section \ref{CPS}. \textit{First}, the \textit{social cost} $J$ is computed as in Section \ref{CoDES} using the CoDES algorithm. \textit{Second}, we find the disagreement point as follows. Each agent computes its own \textit{selfish cost} $D_i$ by ignoring all other players and optimizing its power trading with the grid based on its own load profile, energy storage parameters, and renewable generation resources. For the $i$th agent, denote the power drawn from and injected into the grid as $P_{i,G}^ + \left( t \right)$ and $P_{i,G}^ - \left( t \right)$, respectively. Then the following constraints must be satisfied:
\begin{equation}
\label{eq:ith_PG+_cons}
0 \leqslant P_{i,G}^ + \left( t \right) \leqslant P_G^{\max }, \quad 0 \leqslant P_{i,G}^ - \left( t \right) \leqslant P_G^{\max },
\end{equation}
and $P_{i,G}(t)=P_{i,G}^+(t)-P_{i,G}^-(t)$. Similarly to (\ref{eq:Power_balance}), the power balance for agent $i=1,\cdots,r$ must be satisfied as:
\begin{equation}
\label{eq:ith_PowerBalance}
{P_{i,D}}\left( t \right) = {P_{i,B}}\left( t \right) + {P_{i,R}}\left( t \right) + \left( {P_{i,G}^ + \left( t \right) - P_{i,G}^ - \left( t \right)} \right).
\end{equation}
Finally, (\ref{eq:Ba_cons}) must hold, and (\ref{eq:ith_PG+_cons}) and (\ref{eq:ith_PowerBalance}) also apply to passive users (e.g. Agent 2 in Fig. \ref{fig:sym}) with $P_{i,B}(t)=P_{i,R}(t)=0$.

Combine all power rating commands of the $i$th agent from time step 1 to $T$ into a vector ${{\mathbf{P}}_i^\mathbf{s}} \in {{\mathbb{R}}^{3T \times 1}}$:
\begin{equation}
\label{eq:P}
{{\mathbf{P}}_i^\mathbf{s}} = {[\begin{array}{*{20}{c}}
  {{\mathbf{P}}_{i,G}^ + }&{{\mathbf{P}}_{i,G}^ - }&{{{\mathbf{P}}_{i,B}}} 
\end{array}]'}
\end{equation}
where 
\begin{equation}
\begin{gathered}
  {\mathbf{P}}_{i,G}^ +  = \left[ {\begin{array}{*{20}{c}}
  {P_{i,G}^ + \left( 1 \right)}& \cdots &{P_{i,G}^ + \left( T \right)} 
\end{array}} \right], \hfill \\
  {\mathbf{P}}_{i,G}^ -  = \left[ {\begin{array}{*{20}{c}}
  {P_{i,G}^ - \left( 1 \right)}& \cdots &{P_{i,G}^ - \left( T \right)} 
\end{array}} \right], \hfill \\
  {{\mathbf{P}}_{i,B}} = \left[ {\begin{array}{*{20}{c}}
  {{P_{i,B}}\left( 1 \right)}& \cdots &{{P_{i,B}}\left( T \right)} 
\end{array}} \right]. \hfill \\ 
\end{gathered}
\end{equation}

The \textit{selfish} cost of agent $i$ is calculated as:
\begin{equation}
\label{eq:min_Ci}
{D_i}=\mathop {\min }\limits_{{\mathbf{P}}_i^\mathbf{s}} \sum\nolimits_{t = 1}^T {\left( {{p_b}\left( t \right)P_{i,G}^ + \left( t \right) - {p_s}\left( t \right)P_{i,G}^ - \left( t \right)} \right)} \Delta t.
\end{equation}
A positive $D_i$ indicates the largest cost agent $i$ is willing to pay for purchasing power from the main grid. If $D_i$ is negative, the active agent $i$ expects at least the profit $-D_i$ by selling its generated power to the main grid.

For the upcoming time steps $1,\cdots, T$, define the price vector ${\mathbf{p}} \in {{\mathbb{R}}^{1 \times 2T}}$ and the demand and renewable generation vectors of agent $i$ as ${{\mathbf{P}}_{i,D}} \in {{\mathbb{R}}^{1 \times T}}$ and ${{\mathbf{P}}_{i,R}} \in {{\mathbb{R}}^{1 \times T}}$, respectively:
\vspace{-0.05in}
\begin{equation}
\resizebox{.89\hsize}{!}{$
\begin{gathered}
  {\mathbf{p}} = \left[ {\begin{array}{*{20}{c}}
  {{p_{b}}\left( 1 \right)}& \cdots &{{p_{b}}\left( T \right)} 
\end{array}|\begin{array}{*{20}{c}}
  {-{p_{s}}\left( 1 \right)}& \cdots &{-{p_{s}}\left( T \right)} 
\end{array}} \right], \hfill \\
 {{\mathbf{P}}_{i,D}} = \left[ {\begin{array}{*{20}{c}}
  {{P_{i,D}}\left( 1 \right)}& \cdots &{{P_{i,D}}\left( T \right)} 
\end{array}} \right], \hfill \\
  {{\mathbf{P}}_{i,R}} = \left[ {\begin{array}{*{20}{c}}
  {{P_{i,R}}\left( 1 \right)}& \cdots &{{P_{i,R}}\left( T \right)} 
\end{array}} \right]. \hfill \\ 
\end{gathered} $}
\end{equation}
Denote vector ${\mathbf{f}} = \left[ {\begin{array}{*{20}{c}}
  {\mathbf{p}}&{{{\mathbf{0}}^{1 \times T}}} 
\end{array}} \right]'$. Then the optimization (\ref{eq:min_Ci}) can be expressed as a linear programming (LP) problem:
\begin{eqnarray}
\label{eq:LPP_mincost}
&&  \mathop {\min }\limits_{{\mathbf{P}}_i^\mathbf{s}} {{\mathbf{f}}'}{{\mathbf{P}}_i^\mathbf{s}} \hfill \\
  s.t.\; &&  {\mathbf{A}}{{\mathbf{P}}_i^\mathbf{s}} \leqslant {\mathbf{B}},  \quad
   {{\mathbf{A}}_{{\mathbf{eq}}}}{{\mathbf{P^s}}_i} = {{\mathbf{B}}_{{\mathbf{eq}}}}, \nonumber \hfill
\end{eqnarray}
where $\mathbf{A}$, $\mathbf{B}$, ${\mathbf{A}}_{{\mathbf{eq}}}$ and ${\mathbf{B}}_{{\mathbf{eq}}}$ are defined by the inequality and equality constraints in (\ref{eq:Ba_cons}), (\ref{eq:ith_PG+_cons}), (\ref{eq:ith_PowerBalance}) and Table \ref{tab:notation}.

By solving the above LP problem, each agent computes its own selfish cost ${D_i} = {{\mathbf{f}}'}{{\mathbf{P}}_i^\mathbf{s}}^*$. The allocated cost of agent $i$ is given by (\ref{eq:Ji}). Finally, note the set of selfish power schedules computed using (\ref{eq:min_Ci}) for all agents is a suboptimal solution of the social optimization (\ref{eq:min_cost_linear}), i.e. (\ref{eq:JlessD}) must hold. Thus, successful bargaining is guaranteed in our game.

\subsection{Distributed Cost Allocation}\label{DNBS}
In (\ref{eq:Ji}), each agent needs the knowledge of other players' selfish costs to compute the sum $\sum\nolimits_{i = 1}^{r} {{D_i}}$. To avoid exchange of these costs, we employ the averaging consensus algorithm \cite{4118472} where the agents and the grid communicate only with their neighbors. Set the initial state of agent $i$ to $x_i(0)=D_i$, and set the grid's initial state to $x_{r+1}(0)=-J$. The state update equation of each node $i=1,\cdots, r+1$ is given by:
\begin{equation}
\label{eq:xi_up}
{x_i}(k + 1) = {x_i}(k) + \sum\nolimits_{j \in {N_i}}^{} {{\omega_{ij}}\left( {{x_j}(k) - {x_i}(k)} \right)},
\end{equation}
Provided the communication network is connected, the state information $x_i(k)$ converges to the average of the initial states \cite{4118472}: 
\begin{equation}
\label{eq:x_avg}
{{\hat x}_i} = \mathop {\lim }\limits_{k \to \infty } {x_i}(k) = \frac{\sum\nolimits_{i = 1}^{r + 1} {{x_i}(0)}}{{r + 1}}  = \frac{{\sum\nolimits_{i = 1}^r {{D_i}}  - J}}{{r + 1}}.
\end{equation}

Thus, at convergence, each agent is able to calculate its own allocated cost $J_i$ using only its local information:
\begin{equation}
\label{eq:Ji_dis}
{J_i} = {D_i} - \frac{{{(r+1)}{{\hat x_i}}}}{{r}}, \quad \forall i = 1, \cdots ,r
\end{equation}
which is equivalent to (\ref{eq:Ji}).

\begin{table}[h!]
\vspace{-0.2in}
\caption{DESD Parameters}
\vspace{-0.1in}
\centering
\begin{tabular}{|c|c|c|c|c|}
\hline
Agent & $E_{i,B}^0$ & $E_{i,B}^{min}$ & $E_{i,B}^{max}$ & $P_{i,B}^{max}$ \\
\hline
Agent 1 & 2.8kWh & 2.8kWh & 7kWh & 3.3kW\\
\hline
Agent 3 & 2.8kWh & 2.8kWh & 10kWh & 4.3kW\\
\hline
\end{tabular}
\label{tab:DESD_param}
\vspace{-0.1in}
\end{table}

\section{Numerical Simulations}

\begin{figure}[!t]
\centering
\includegraphics[width=0.48\textwidth]{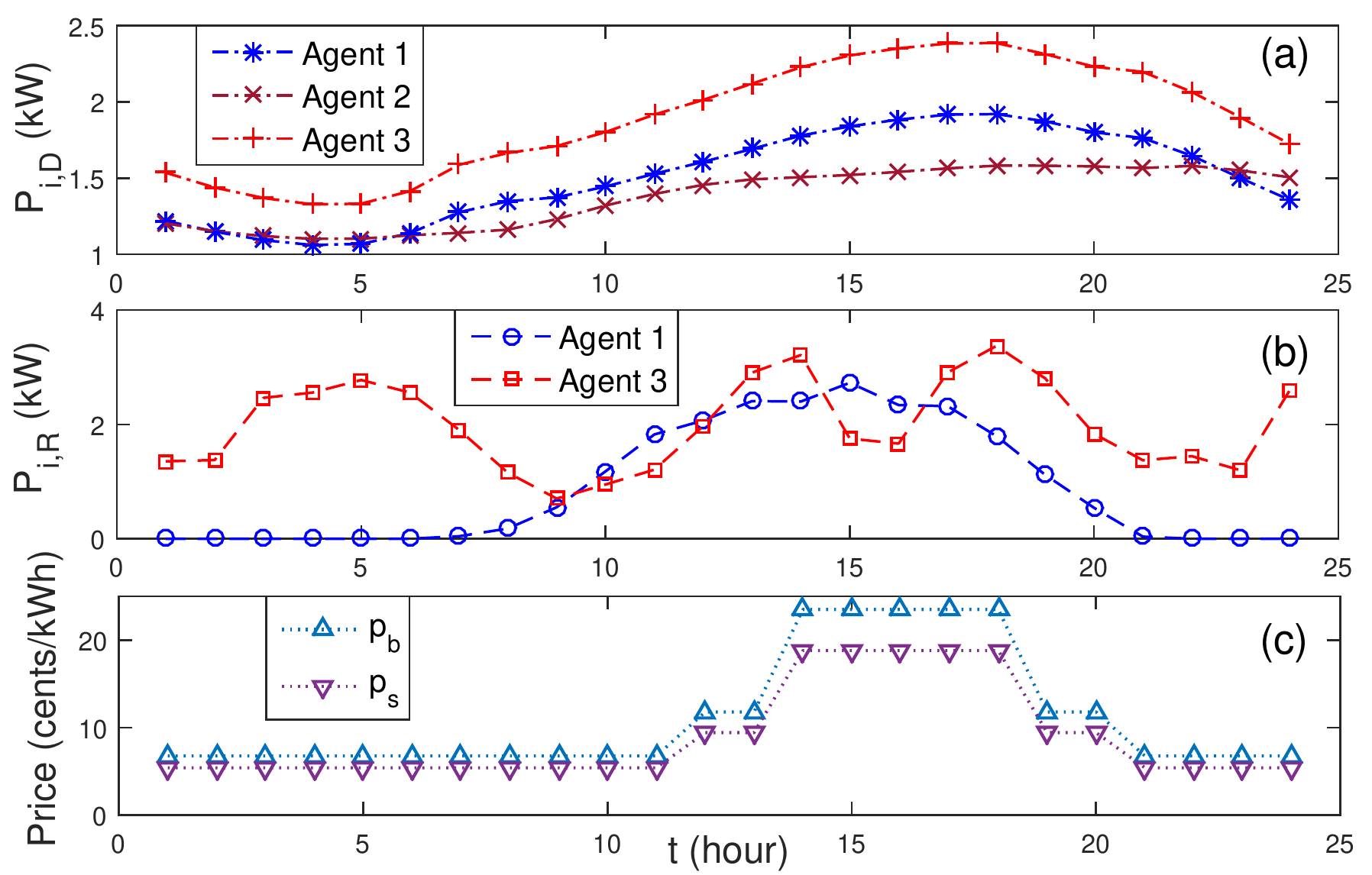}
\vspace{-0.1in}
\caption{System parameters over the 24 hours period. (a) Users' power demands; (b) Active users' renewable generation; (c) The price profile.}
\label{fig:profile}
\vspace{-0.15in}
\end{figure}

We use the 4-Bus microgrid system model (3 agents and 1 grid) shown in Fig. \ref{fig:sym} to validate the extended CoDES method in Section \ref{CPS} and the cost allocation approach in Section \ref{NBS}. Each agent owns a distributed controller, which executes the proposed algorithms and exchanges state information with its neighbors' controllers via bidirectional communication links. To protect the users' data, the local information of each Bus, including its renewable generation and power consumption, is only accessible to the controller embedded in that Bus. The demand profiles of all users and the renewable generation of the active users are obtained from PJM database\footnote{http://pjm.com/markets-and-operations/energy/real-time/hourly-prelim-loads.aspx}. The price $p_b(t)$ is obtained from the Duke Energy Progress, NC residential service schedule tariff\footnote{https://www.duke-energy.com/pdfs/R3-NC-Schedule-R-TOU-dep.pdf}, and $p_s(t)$ is set to $80\%$ of $p_b(t)$. The DESD parameters are listed in Table \ref{tab:DESD_param}. The demand and renewable generation of the agents are shown in Fig. \ref{fig:profile}(a,b), and the price profile is illustrated in Fig. \ref{fig:profile}(c). The interval $T=24$ steps, and $\Delta t=1$ hr.

\begin{figure}[!t]
\centering
\includegraphics[width=0.47\textwidth]{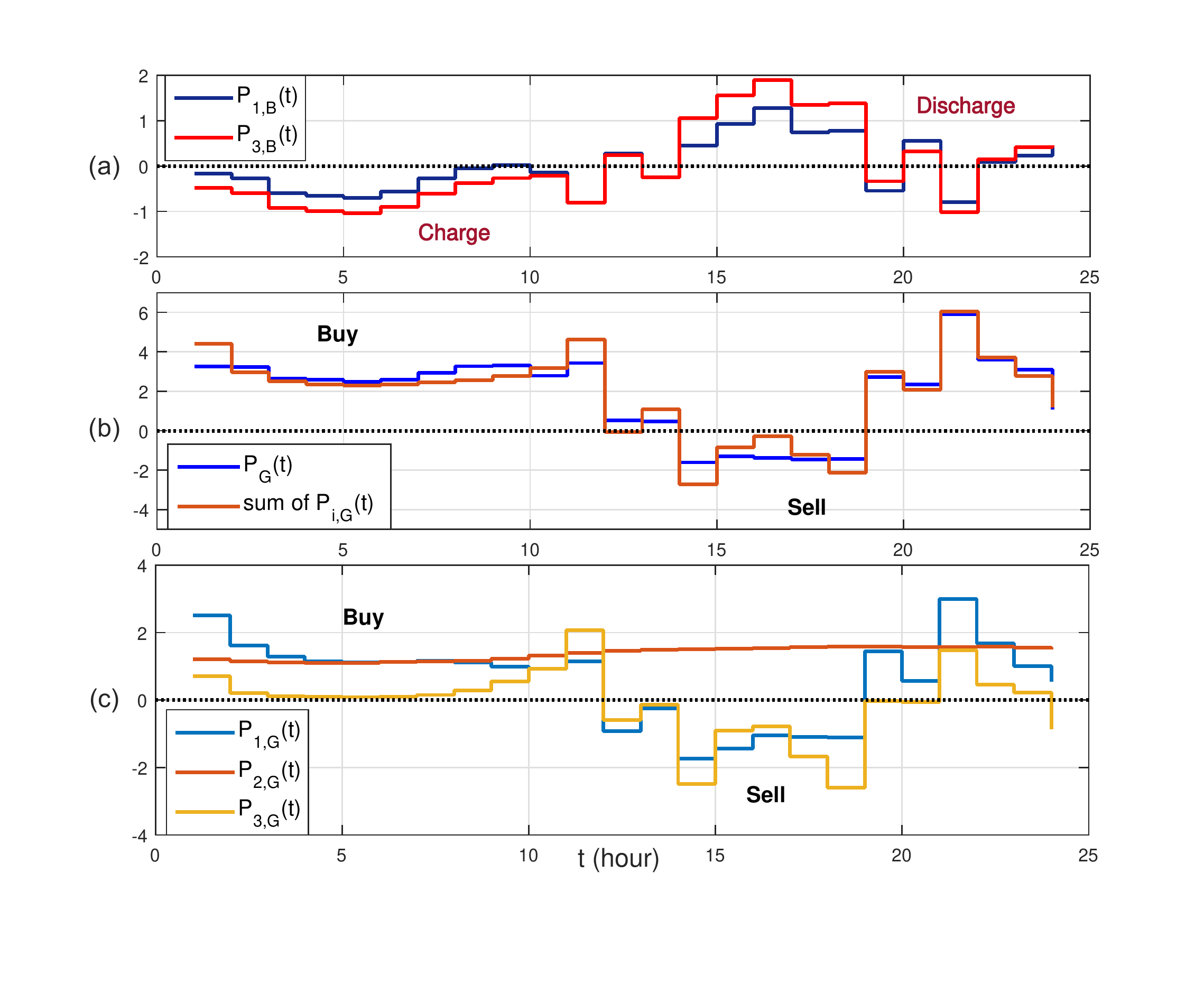}
\vspace{-0.1in}
\caption{Optimal day-ahead power schedules. (a) Power commands to DESDs of active agents using the extended \textit{CoDES} algorithm; (b) Grid power schedule using the extended \textit{CoDES} and the sum of all agents' powers traded with the grid at the disagreement point $\mathbf{D}$ (\ref{eq:min_Ci}); (c) Agents' powers traded with the grid at $\mathbf{D}$.}
\label{fig:sche}
\vspace{-0.2in}
\end{figure}

Fig. \ref{fig:sche} shows the optimal day-ahead schedules for the proposed algorithms. In Fig. \ref{fig:sche}(a), the power commands to the DESDs of the active users $i$=1,3 are illustrated for the CoDES method while Fig. \ref{fig:sche}(b) shows the schedules of the power traded with the grid for social optimization (\ref{eq:min_cost_linear}) and of the combined powers of the selfish optimizations (\ref{eq:min_Ci}) of all agents. Note that during the off-peak hours (1-11 hr, 21-24 hr), when the price is relatively low, the active users tend to buy power and charge the DESDs while they use the stored power (discharge) or sell surplus generated power to the grid during the peak hours (14-18 hr), when the utility price is higher. Moreover, Fig. \ref{fig:sche}(c) reveals significant differences in users' power profiles that result from stand-alone power trading with the grid in (\ref{eq:min_Ci}). We observe that the passive $2^{\mathrm{nd}}$ agent buys power at a relatively constant rate to satisfy its demand while the active users adapt to price variation by utilizing their storage devices and renewable generation. In particular, the $3^{\mathrm{rd}}$ agent sells more power than it buys from the grid due to its active renewable generation and large storage capacity (Table \ref{tab:DESD_param}), despite its high power demand (Fig. \ref{fig:profile}(a)).

\begin{figure}[!t]
\centering
\includegraphics[width=0.48\textwidth]{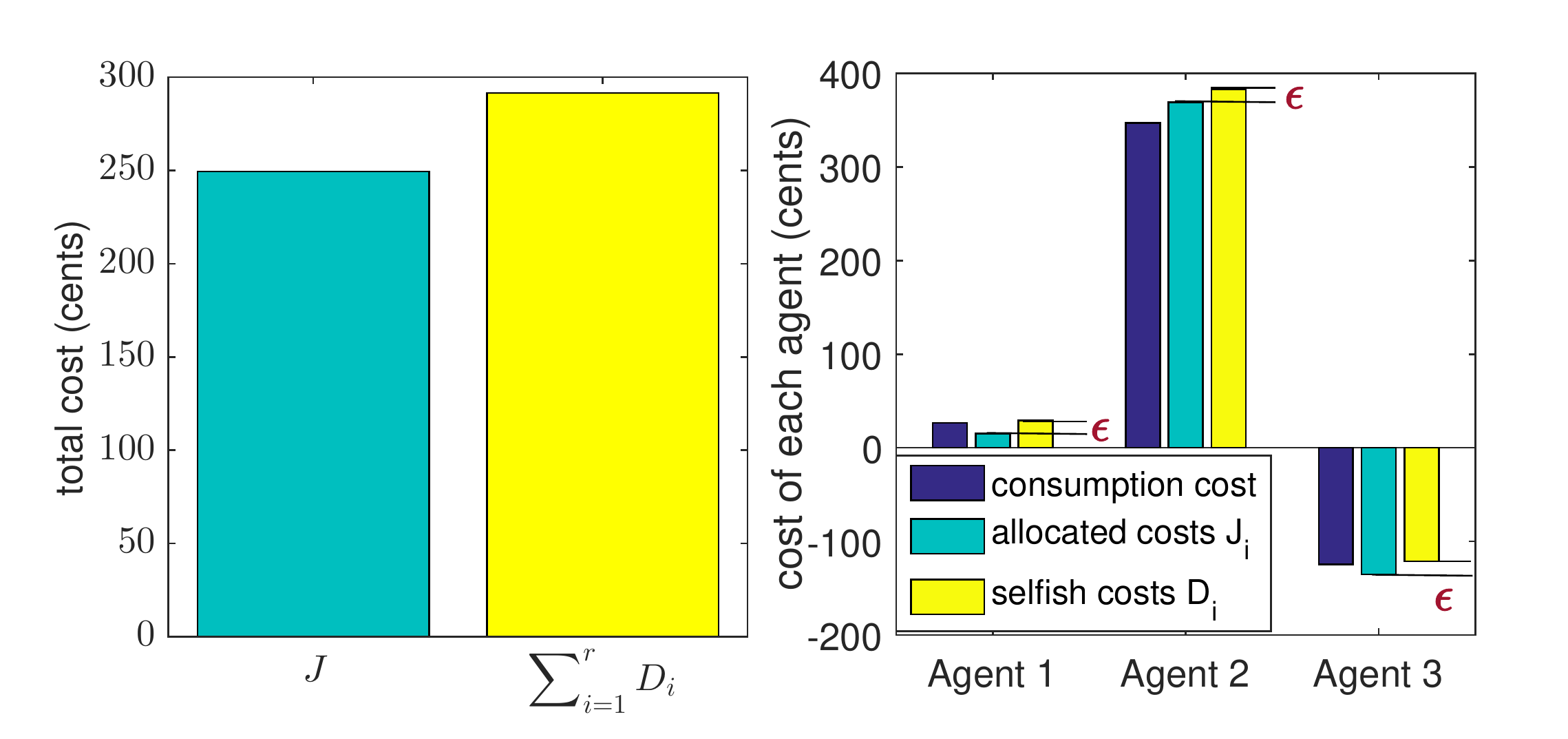}
\vspace{-0.1in}
\caption{Cost allocation. (a) The social cost $J$ (\ref{eq:min_cost_linear}) and the sum of selfish costs $D_i$; (b) selfish, allocated, and consumption costs.}
\label{fig:cost}
\vspace{-0.2in}
\end{figure}

In Fig. \ref{fig:cost}, results of the cost allocation algorithm in Section \ref{cl} are illustrated. From Fig. \ref{fig:cost}(a), we observe that eq. (\ref{eq:JlessD}) holds, confirming successful bargaining and savings of about $45 \cent$ per day due to cooperation. In Fig. \ref{fig:cost}(b), we show the selfish, allocated, and consumption costs of individual agents. We observe that allocated costs $J_i$ provide a fixed discount $\epsilon \approx 15 \cent$ (\ref{eq:eps}) per user per day relative to the selfish costs $D_i$ (\ref{eq:min_Ci}). Moreover, Agent 2 pays the most while Agent 3 is compensated for selling its generated power to the grid, which is consistent with its selfish power scheduling in Fig. \ref{fig:sche}(c). In addition, it receives a cooperation profit $\epsilon$ by providing its stored power to other users.


Finally, we found that the CoDES algorithm in Section \ref{CoDES} converged within 2000 iterations (6 seconds\footnote{The experiments are run using MATLAB on a Dell OPTIPLEX 980 with Windows 7 64-bit operating system, 2.93GHz i7 CPU and 8GB RAM.}) to the schedule found by the optimal centralized algorithm for solving (\ref{eq:min_cost_linear}) using linear programming, which was implemented using the MATLAB built-in function \textit{linprog.m}. Moreover, the distributed NBS method in Section \ref{DNBS} converged in 10 iterations ($<$ 1 sec). The latter consensus-based algorithm converges faster than the social optimization method because its only purpose is to find the average (\ref{eq:xi_up}) while the CoDES algorithm also optimizes the power schedules of the agents.

\section{Conclusion}
Distributed cost optimization and allocation was investigated for microgrids that contain energy storage and renewable generation devices. First, we developed a fully-distributed multi-step energy scheduling method that optimizes both purchasing power from and selling surplus power to the main grid. Second, we proposed a distributed NBS-based cost allocation methods. Finally, we demonstrated practical relevance and fast convergence of proposed algorithms using a microgrid model with heterogeneous loads, activity levels, and time-varying power prices.


%
%

\bibliographystyle{ieeetr}
\bibliography{ref}

\end{document}